\documentclass[aps,twocolumn,pra,groupedaddress,superscriptaddress,nofootinbib]{revtex4-1}

\usepackage{graphicx,amsmath,amssymb,amsbsy,subfigure,hyperref,bbm,times,txfonts,float}
\usepackage{subfigure,hyperref,bbm,times}
\usepackage[T1]{fontenc}
\usepackage{braket}
\usepackage{epsfig} 
\usepackage{color}
\usepackage{graphicx}% Include figure files
\usepackage{dcolumn}
\usepackage{bm}% bold math

\providecommand{\openone}{\leavevmode\hbox{\small1\kern-3.8pt\normalsize1}}

\usepackage{soul}

\hypersetup{
	colorlinks=true,
	linkcolor=blue,
}
\graphicspath{{figure/}}

\begin{document}
	
	%\preprint{APS/123-QED}
	
	\title{Memory effects in high-dimensional systems faithfully identified by Hilbert-Schmidt speed-based witness}% Force line breaks with \\
	
\author{Kobra Mahdavipour}%
\affiliation{Dipartimento di Ingegneria, Universit\`{a} di Palermo, Viale delle Scienze,  90128 Palermo, Italy}%
\affiliation{INRS-EMT, 1650 Boulevard Lionel-Boulet, Varennes, Qu\'{e}bec J3X 1S2, Canada}

\author{Mahshid Khazaei Shadfar}%
\affiliation{Dipartimento di Ingegneria, Universit\`{a} di Palermo, Viale delle Scienze,  90128 Palermo, Italy}%
\affiliation{INRS-EMT, 1650 Boulevard Lionel-Boulet, Varennes, Qu\'{e}bec J3X 1S2, Canada}

	\author{Hossein Rangani Jahromi}
\email{h.ranganijahromi@jahromu.ac.ir}%
\affiliation{Physics Department, Faculty of Sciences, Jahrom University, P.B. 74135111, Jahrom, Iran}

\author{Roberto Morandotti}
	\affiliation{INRS-EMT, 1650 Boulevard Lionel-Boulet, Varennes, Qu\'{e}bec J3X 1S2, Canada}

\author{Rosario Lo Franco}%
 \email{rosario.lofranco@unipa.it}
\affiliation{Dipartimento di Ingegneria, Universit\`{a} di Palermo, Viale delle Scienze,  90128 Palermo, Italy}%

	\date{\today}% It is always \today, today,
	
	\begin{abstract}
		A witness of non-Markovianity based on the Hilbert-Schmidt speed (HSS), a special type of quantum statistical speed, has been recently introduced for low-dimensional quantum systems.  Such a non-Markovianity witness is particularly useful, being easily computable since no diagonalization  of the system density matrix  is required. We investigate the sensitivity of this HSS-based witness to detect non-Markovianity in various high-dimensional and multipartite open quantum systems. We find that the time behaviors of the HSS-based witness are always in agreement with those of quantum negativity or quantum correlation measure. These results show that the HSS-based witness is a faithful identifier of the memory effects appearing in the quantum evolution of a high-dimensional system.
	\end{abstract}
	
	%\keywords{Non-Markovian dynamics, Hilbert-Schmidt speed.}
	%Use showkeys class option if keyword
	
	
	\maketitle

	The unavoidable interaction of quantum systems with their environments induces decoherence and dissipation of energy.  Recently, because of important developments
in both theoretical and experimental branches of quantum information theory, studies of memory effects (non-Markovianity) during the evolution of quantum systems have attracted much attention (see Refs.~\cite{rivas2014quantum,deVegaReview,BreuerColloquium} for some reviews). Some approaches used for a quantitative description of non-Markovian processes are either related to the presence of information backflows \cite{PhysRevLett.103.210401} or to the indivisibility of the dynamical map \cite{PhysRevLett.105.050403}. However, while well-defined for classical evolution, the notion of non-Markovianity appears to still lack a unique definition in the quantum scenario \cite{banacki2020information}. 

Non-Markovian processes, exhibiting quantum memory effects, have been characterized and observed in various realistic systems such as quantum optical systems \cite{Tang_2012,LiuSciRep,Liu2011,XuNatComm2013,MataloniSciRep,orieux2015}, superconducting qubits \cite{Modi2020,PhysRevX.11.041043}, photonic crystals \cite{PhysRevA.78.060302,PhysRevLett.108.043603,Burgess:21}, light-harvesting complexes \cite{Chin_2010}, chemical compounds \cite{Shao_2012,Pomyalov_2005}. Moreover, it is known that non-Markovianity can be a resource for quantum information tasks \cite{PhysRevLett.108.160402,PhysRevA.83.042321,PhysRevLett.109.233601,laine2014nonlocal,dong2018non}. Accordingly, various witnesses have been proposed to identify non-Markovianity based on, for example, distinguishability between evolved quantum states of the system \cite{PhysRevLett.103.210401}, fidelity \cite{PhysRevA.84.052118,PhysRevA.82.042107,jahromi2019multiparameter}, quantum relative entropies \cite{PhysRevA.81.062115,PhysRevA.83.022109}, quantum Fisher information \cite{PhysRevA.82.042103}, capacity measure \cite{bylicka2013non,PhysRevA.89.012114,PhysRevA.89.024101} and Bloch volume measure \cite{PhysRevA.88.020102,PhysRevA.90.012113,PhysRevA.90.012310}.

%In recent years, the vast majority attention has been paid to
%the characterization and quantification of non-Markovianity. One of the  well-known witness proposed by Breuer-Laine-Piilo $(BLP)$ \cite{PhysRevLett.103.210401} which is based on distinguishability (trace distance  \cite{PhysRevLett.103.210401}) between quantum
%states of the same systems. Other well-established non-markovianity quantifier relies on dynamical divisibility. The
%nondivisibility of a completely positive and
%trace preserving (CPTP) \cite{PhysRevLett.105.050403} maps is necessary for the occurrence
%of information back-flow from the environment or the presence
%of the environment memory \cite{PhysRevLett.103.210401}. Mainly form that physical point of view, 

It has been shown that the nonmontonic behavior of quantum resources such as entanglement  \cite{Rivas_2014}, quantum coherence \cite{PhysRevLett.113.140401,PhysRevLett.116.120404,PhysRevLett.116.070402,RevModPhys.89.041003} and quantum mutual information \cite{PhysRevA.86.044101} can be interpreted as a witness of
quantum non-Markovianity.  Using entanglement to witness non-Markovianity
was first proposed in Ref.~\cite{PhysRevLett.105.050403}. This proposal has been theoretically
investigated for qubits coupled to bosonic environments \cite{PhysRevA.84.032118,PhysRevA.85.052104,PhysRevA.84.032124}, for a damped harmonic
oscillator \cite{PhysRevA.89.022109}, and for random unitary dynamics and classical noise models \cite{PhysRevA.85.032318}. It is also shown that entanglement cannot capture all the quantumness of correlations because there are some separable mixed states with vanishing
entanglement, while they can have nonzero quantum correlations \cite{PhysRevLett.100.050502}. Therefore, in this sense, quantum correlations are more robust than entanglement  \cite{PhysRevLett.107.170502, PhysRevA.85.010102, PhysRevA.80.044102, bellomo2011dynamics}, while entanglement may suffer sudden death \cite{yu2009sudden,almeida2007environment}. Consequently, many methods to quantify quantum correlations have been provided, among which quantum discord \cite{ollivier2001quantum,henderson2001classical} and measurement-induced disturbance \cite{PhysRevA.77.022301} are proper for any bipartite state.

%In more investigation for the non-Markovian reservoir, the dynamics of a pair of two level systems in an initial Bell-like state \cite{PhysRevLett.99.160502}, Werner-like state \cite{PhysRevA.77.032342}, or the dynamics of two
%oscillators \cite{PhysRevLett.100.220401,PhysRevA.79.032102}, the spin-S systems embedded in dephasing environments \cite{Fan_2014} may show the
%presence of entanglement oscillations and revivals after a
%finite period of time of its complete disappearance. 

Recently, Hilbert-Schmidt speed (HSS) \cite{gessner2018statistical}, a measure of
quantum statistical speed which has the advantage of avoiding diagonalization of the evolved density matrix, has been proposed and employed as a faithful witness of non-Markovianity in Hermitian systems \cite{PhysRevA.102.022221,RanganiJahromi2021,jahromi2021witnessing,HosseinPLA2022}.
These studies are so far especially limited to low-dimensional systems, while high-dimensional ones have not been investigated in detail. We know that high-dimensional systems play a crucial role in increasing the security in quantum cryptography \cite{PhysRevLett.88.127901,PhysRevLett.88.127902}, as well as in enhancing quantum logic gates, fault-tolerant quantum computation and quantum error correction \cite{10.1007/3-540-49208-9_27}.
This motivates us to check the sensitivity of HSS-based witness to detect non-Markovianity in high-dimensional and multipartite open quantum systems. 

In this work, we analyze the validity of our HSS-based witness in various examples of high-dimensional open quantum systems, such as qudits and hybrid qubit-qutrit systems. In particular, we consider a single qudit (spin-S systems) subject to a squeezed vacuum reservoir \cite{JI2020164088}, and hybrid qubit-qutrit system coupled to quantum as well as classical noises \cite{TCHOFFO20191856}. We observe that the HSS-based witness is consistent with established non-Markovianity quantifiers based on dynamical breakdown of monotonicity for the quantum information resources.

The paper is organized as follows. In Sec.~\ref{II} we briefly review the definition of quantifiers. In Sec.~\ref{III} the  sensitivity of HSS-based witness in high-dimensional and multipartite open quantum systems through various examples is studied.  Finally, Sec.~\ref{IV} summarizes the main results and prospects.

%%%%%%%%%%%%%%%%%%%%%%%%%%%%%%%%%%%%%%%%%%
\section{Definition of the quantifiers}\label{II}
In this section we briefly review the relevant quantifiers employed in this paper.

\subsection{HSS-based witness of non-Markovianity}
Considering the distance measure \cite{PhysRevA.97.022109}
\begin{equation}
	[\text{d}(p,q)]^{2}=\dfrac{1}{2}\sum\limits_{x}^{}|p_{x}-q_{x}|^{2},
\end{equation}
where $ p = \{p_{x}\}_{x} $ and $ q = \{q_{x}\}_{x} $ denote the probability distributions, one   can quantify the distance between
infinitesimally close distributions taken from a one-parameter
family $ p_{x} (\phi) $ and then
 define  the classical statistical speed as
\begin{equation}\label{classicalspeed}
	\text{s}\big[p(\phi_{0})\big]=\dfrac{d}{d\phi}\text{d}\big(p(\phi_{0}+\phi),p(\phi_{0})\big).
\end{equation}
These classical notions can be generalized  to the quantum case  by
assuming a  pair of quantum states $ \rho $ and $ \sigma $, and writing  $ p_{x} = \text{Tr}\{E_{x}\rho\} $ and $ q_{x} = \text{Tr}\{E_{x}\sigma\} $ which represent the measurement
probabilities corresponding to the positive-operator-valued measure (POVM) defined by the $ \{E_{x}\geq 0\} $ satisfying $\sum\limits_{x}^{} E_{x} = \mathbb{I} $.
 The associated quantum distance which called Hilbert-Schmidt distance \cite{OZAWA2000158} can be achieved by maximizing  the classical distance over all  possible choices of
POVMs \cite{PhysRevA.69.032106}
\begin{equation}\label{qdis}
	\text{D}(\rho,\sigma)\equiv\max_{\{E_{x}\}}\text{d}(p,q)=\sqrt{\frac{1}{2}\text{Tr}
		{\left[\left(\rho-\sigma\right)^{2}\right]}}.
\end{equation}
Consequently, the HSS, the corresponding quantum statistical speed is defined as follows
\begin{equation}\label{quantumspeed}
	HS\!S\left(\rho_{\phi}\right)\equiv HS\!S_{\phi}\equiv\max_{\{E_{x}\}} \text{s}\big[p(\phi)\big]=\sqrt{\frac{1}{2}\text{Tr}\left[\bigg(\dfrac{d\rho_{\phi}}{d\phi}\bigg)^{2}\right]},
\end{equation}
which can be easily computed without diagonalization of $ \dfrac{\text{d}\rho_{\phi}}{\text{d}\phi} $.

\par
Now  the recently proposed protocol
to detect the non-Markovianity based on
the HSS \cite{PhysRevA.97.022109} is briefly recalled. We consider an $n$-dimensional quantum system whose initial state is given by
\begin{equation}\label{eq:Eq5}
	|\psi_{0}\rangle=\dfrac{1}{\sqrt{n}}\big(\text{e}^{i\phi}|\psi_{1}\rangle+\ldots+|\psi_{n}\rangle\big),
\end{equation}
where $\phi$ is an unknown phase shift and $\{|\psi_{1}\rangle,...,|\psi_{n}\rangle\} $ denotes a complete and orthonormal set (basis) for the corresponding Hilbert space $ \mathcal{H} $. Given this initial state, the HSS-based witness of non-Markovianity is defined by
\begin{equation}\label{eq:Eq6}
	\text{Non-Markovianity Witness}:\chi(t)\equiv \dfrac{\text{d}HS\!S \big(\rho_{\phi}(t)\big)}{\text{d}t} > 0,
\end{equation}
in which $ \rho_{\phi}(t) $ is the evolved state of the system.

\subsection{Quantum entanglement measure}
Quantum entanglement is a kind of quantum correlations which, from an operational point of view, can be defined as those correlations between different subsystems which cannot be generated by local operations and classical communication (LOCC) procedures. We use negativity  \cite{plenio2005logarithmic} to quantify the quantum entanglement of the state, which is a reliable measure of entanglement in the case of qubit-qubit and qubit-qutrit systems \cite{nakahara2008quantum}.

 For any bipartite state $\rho_{AB}$ the negativity is defined as     
\begin{equation}
	\mathcal{N}{\left(\rho_{AB}\right)}=\sum_{i}\vert \lambda_{i}\vert,
\end{equation}
where $\lambda_{i}$ is the negative eigenvalue of $\rho^{T_{k}}$, with $\rho^{T_{k}}$ denoting the
partial transpose of the density matrix $\rho_{AB}$ with
respect to subsystem $k = A, B$. The negativity can also be computed by the following formula \cite{jaeger2007quantum}:
\begin{equation}
	\mathcal{N}{\left(\rho_{AB}\right)}=\frac{1}{2}{\left(\left\|{\rho^{T_{k}}}\right\|-1\right)},
\end{equation}
in which the trace norm of  $\rho^{T_{k}}$ is equal to the sum of the absolute
values of its eigenvalues  \cite{wilde2013quantum}:
\begin{equation}
	\left\|{\rho^{T_{k}}}\right\|=\sum_{i}\vert \mu_{i}\vert,
\end{equation}
 where the spectral decomposition of $ \rho^{T_{k}} $ is given by $ \sum_{i}\mu_{i} \ket{i}\bra{i}$.  

\subsection{Quantum correlation quantifier: Measurement-induced disturbance}
We use measurement-induced disturbance MID \cite{luo2008using} as an alternative nonclassicality indicator for quantifying the quantum correlations of the bipartite quantum 
systems. It is defined as the minimum disturbance
caused by local projective measurements  leaving the reduced states invariant.

Considering the spectral resolutions of the reduced density states $ \rho_{A}=\sum_{i}^{}p_{i}^{A}\Pi_{i}^{A} $ and  $ \rho_{B}=\sum_{j}^{}p_{j}^{B}\Pi_{i}^{B} $, one can compute  the MID as follows
\begin{equation}\label{MIDformul}
	\mathcal{M}{\left(\rho_{AB}\right)}=\mathcal{I}{\rho_{AB}}-\mathcal{I}{\left(\Pi{{\left(\rho_{AB}\right)}}\right)},
\end{equation}
where $ \mathcal{I}$ is the mutual quantum information given by
\begin{equation}
	\mathcal{I}{\left(\rho_{AB}\right)}=S{\left(\rho_{A}\right)}+S{\left(\rho_{B}\right)}-S{\left(\rho_{AB}\right)} ,
\end{equation}
in which $S{\left(\rho\right)}=-\mathrm{tr}{\rho \log{\left(\rho\right)}}$
denotes the von Neumann entropy
and 
\begin{equation}
	\Pi{\left(\rho_{AB}\right)}=\sum_{i,j}\left(\Pi_{i}^{A}\otimes\Pi_{j}^{B}\right)\rho_{AB}\left(\Pi_{i}^{A}\otimes\Pi_{j}^{B}\right).
\end{equation}

%%%%%%%%%%%%%%%%%%%%%%%%%%%%%%%%%%%%%%%%%%
\section{Analyzing the efficiency of the HSS witness in high-dimensional systems}\label{III}
In this section we check the sanity of HSS-based witness through several paradigmatic high dimensional quantum systems. The analyses are based on the fact that  the revivals of 
 quantum correlations  are associated with the non-Markovian evolution of the system \cite{PhysRevA.85.032318}.
 In particular, we consider a single qudit subject to a quantum environment and  a hybrid qubit-qutrit system coupled to  independent as well as common quantum and classical noises. 
We  show that the oscillation of HSS-based witness is in qualitative agreement with nonmonotonic variations of the quantum resources, and hence it can be introduced as a faithful identifier of non-Markovianity in such high dimensional systems. 
\par
It should be noted that the efficiency of the HSS-based witness in detecting the non-Markovian nature of the dynamics directly depends on  adopting the correct parametrization of the initial state (\ref{eq:Eq5}), as discussed in \cite{PhysRevA.102.022221}. However, often choosing the computational basis as the complete orthonormal set $\{|\psi_{1}\rangle,...,|\psi_{n}\rangle\} $ is enough to capture the non-Markovianity, as shown in this paper. In all examples discussed below, the HSS is computed for the pure initial states while the quantum correlations may be calculated for mixed ones to illustrate the general efficiency off the HSS-based witness.

\subsection{Single-qudit interacting with a quantum environment}\label{Example}
\subsubsection{Coupling to a thermal reservoir}
Let consider the spin-S systems  interacting with a thermal reservoir modeled by an infinite chain of quantum harmonic
oscillators with $\omega_{k}$, $b_{k}$, and $b_{k}^{\dagger}$ being, respectively, the
frequency, annihilation, and creation operators for the $k-th$ oscillator. The total Hamiltonian of the system is given by 
\begin{equation}\label{eq17}
	H=\omega_{0}S_{z}+\sum_{k} \omega_{k} b_{k}^{\dagger} b_{k}+\sum S_{z}(g_{k}b_{k}^{\dagger}+g_{k}^{*}b_{k}),
\end{equation}
in which $\omega_{0}$ denote the transition frequency between any neighboring
energy states of the spin, and $S_{z}$, the $z$ component of spin
operator, can be represented by a diagonal matrix $S_{z}=\mathrm{diag}[s,s-1,\ldots,-s]$ in the eigen-basis $\{\vert i\rangle, i= s, \ldots, -s\}$. In the interaction picture  Eq.~(\ref{eq17}) into is expressed as
\begin{equation}\label{eq:Eq18}
	H_{I}= \sum S_{z}(g_{k}b_{k}^{\dagger}e^{i\omega_{k}t}+g_{k}^{*}b_{k}e^{-i\omega_{k}t}),
\end{equation}
where $g_{k}$ denotes the coupling strength between the spin and the environment through the dephasing interaction.
Up to an overall phase factor, the corresponding unitary propagator is obtained as
\begin{equation}\label{eq:Eq15}
	V{\left(t\right)}=\exp{\left[\frac{1}{2}S_{z}{\sum_{k}{\left(\alpha_{k}b_{k}^{\dagger}-\alpha^{*}b_{k}\right)}}\right]},
\end{equation}
where $\alpha_{k}=2g_{k}\left(1-e^{i\omega_{k}t}\right)/{\omega_{k}}$.  
\par
It is assumed  that the initial state of the spin-bath system is in
a product state $\rho_{T}\left(0\right)=\rho{\left(0\right)}\otimes \rho_{B}$ in which $\rho{\left(0\right)}$ denotes the initial state of spin, and 
\begin{equation}
	\rho_{B}=\frac{1}{Z_{B}}e^{-\beta \sum_{k} \omega_{K}b_{k}^{\dagger}b_{k}}
\end{equation}
represents the thermal equilibrium state of the bath with partition function $Z_{B}$ and inverse temperature $\beta=\frac{1}{k_{B}T}.$
The  evolved state of the system can be calculated by
  \cite{zi2014entanglement}
\begin{equation}\label{eq:Eq21}
	\rho_{nm}(t) =\rho_{nm}(0)\exp{[-(n-m)^2\Gamma(t)]},
\end{equation}
where $n,m=-s,-s+1,\ldots,0,\ldots,s-1, s$ and, in the continuum-mode limit, the decoherence function
is given by
\begin{equation}
	\Gamma(t)=\int_0^{\infty} J(\omega)\coth\left(\frac{\omega}{2k_bT}\right) \frac{1-\cos(\omega t)}{\omega^2} d\omega,
\end{equation}
with spectral density $J{\left(\omega\right)}=\sum_{k}\vert g_{k}\vert^{2}\delta{\left(\omega-\omega_{k}\right)}$.

The   $\Gamma(t)$ behavior  closely depends on the characteristics
of the environment. Here we consider the Ohmic-like reservoirs with spectral density
\begin{equation}
	J(\omega)=\alpha\frac{\omega^s}{\omega_c^{s-1}}\exp\left({\frac{-\omega}{\omega_c}}\right),
\end{equation}
where $\alpha$ represents a dimensionless coupling strength, and $\omega_c$ denotes the cutoff frequency of the bath. Changing the Ohmic parameter
$ s $, one can obtain sub-Ohmic  ($ 0 
< s < 1 $), Ohmic
($ s = 1 $) and super-Ohmic ($ s > 1 $) reservoirs.

\subsubsection{Coupling to a squeezed vacuum reservoir}
In the case that the spin system is coupled to a squeezed vacuum reservoir, the reduced density-matrix elements are similar to ones presented in Eq.~(\ref{eq:Eq21}) when the decoherence function $ \Gamma(t) $ is replaced by
\begin{equation}\label{Squeezdec}
	\begin{split}
		\gamma{(t)}&=\int_0^{\infty} J(\omega)\frac{(1-\cos{\left(\omega t\right)})}{\omega^{2}}
		 {[\cosh{(2r)-\sinh{(2r)\cos{(\omega t-\boldsymbol{\theta})}}}]}d\omega,
	\end{split}
\end{equation}
where $r$ is the squeezed amplitude parameter, and $\boldsymbol{\theta}$ denotes the squeezed angle.

Because the structures of the density matrices are the same in both scenarios (coupling to thermal and  squeezed vacuum reservoirs), we only focus on the interaction of the system with the squeezed vacuum reservoir, noting that the general results also holds for the  
thermal reservoir.

\par
We take the qudit in the pure initial state
\begin{equation}\label{eq:Eq25}
	\vert\psi\rangle=\frac{1}{\sqrt{2s+1}}{(e^{i\phi}{\vert s\rangle +\vert s-1\rangle+\vert s-2\rangle+\cdots+ \vert -s\rangle)}},
\end{equation}
which leads to the evolved state $\rho{\left(t\right)}$ given by
\begin{equation}\label{eq:Eq26}
	\rho{(t)}=\frac{1}{2s+1}\begin{pmatrix}
		1& e^{-\gamma{(t)}}e^{i\phi}&\cdots& e^{-{(2s)^{2}}\gamma{(t)}}e^{i\phi} \\
		e^{-\gamma{(t)}}e^{-i\phi} &1 &  \cdots & e^{-{(2s-1)^{2}}\gamma{(t)}}\\
		e^{-4\gamma{(t)}}e^{-i\phi} &e^{-\gamma{(t)}} & \cdots & e^{-{(2s-2)^{2}}\gamma{(t)}}\\
		\vdots & 1&\ddots\\
		e^{-{(2s)^{2}}\gamma{(t)}}e^{-i\phi}&e^{-{(2s-1)^{2}}\gamma{(t)}}&\cdots & 1
	\end{pmatrix}.
\end{equation}
Therefore, the time derivative of the HSS-based witness is obtained as
\begin{equation}
	\chi{(t)}=-{\frac{1}{2s+1}}\frac{\partial \gamma{(t)}}{\partial t}{\frac{\sum_{k=1}^{2s}{k^{2}}e^{-2k^{2}\gamma{(t)}}}{\sum_{k=1}^{2s}e^{-2k^{2}\gamma{(t)}}}}.
\end{equation}

The HSS-based witness $\chi{(t)}>0$ tells us that the process is non-Markovian whenever $\frac{\partial{\gamma{(t)}}}{\partial t}<0$, which corresponds to time intervals in which the decoherence function decreases, leading to the re-coherence phenomenon. As  known, in this system the non-Markovian effects, originating from the non-divisible maps, appear when the decoherence function  temporarily decays with time \cite{fanchini2017lectures}. Therefore, our witness correctly predicts  the intervals at which the memory effects arise in this single-qudit system.
Moreover, when $\gamma{(t)}$ is a monotonous increasing function of time, the dynamics is Markovian because  the coherence decays
monotonously with time.

\subsection{Hybrid qubit-qutrit system interacting with various quantum and classical environments}
\begin{figure}[H]
	\centering
	\includegraphics[width=0.48\textwidth]{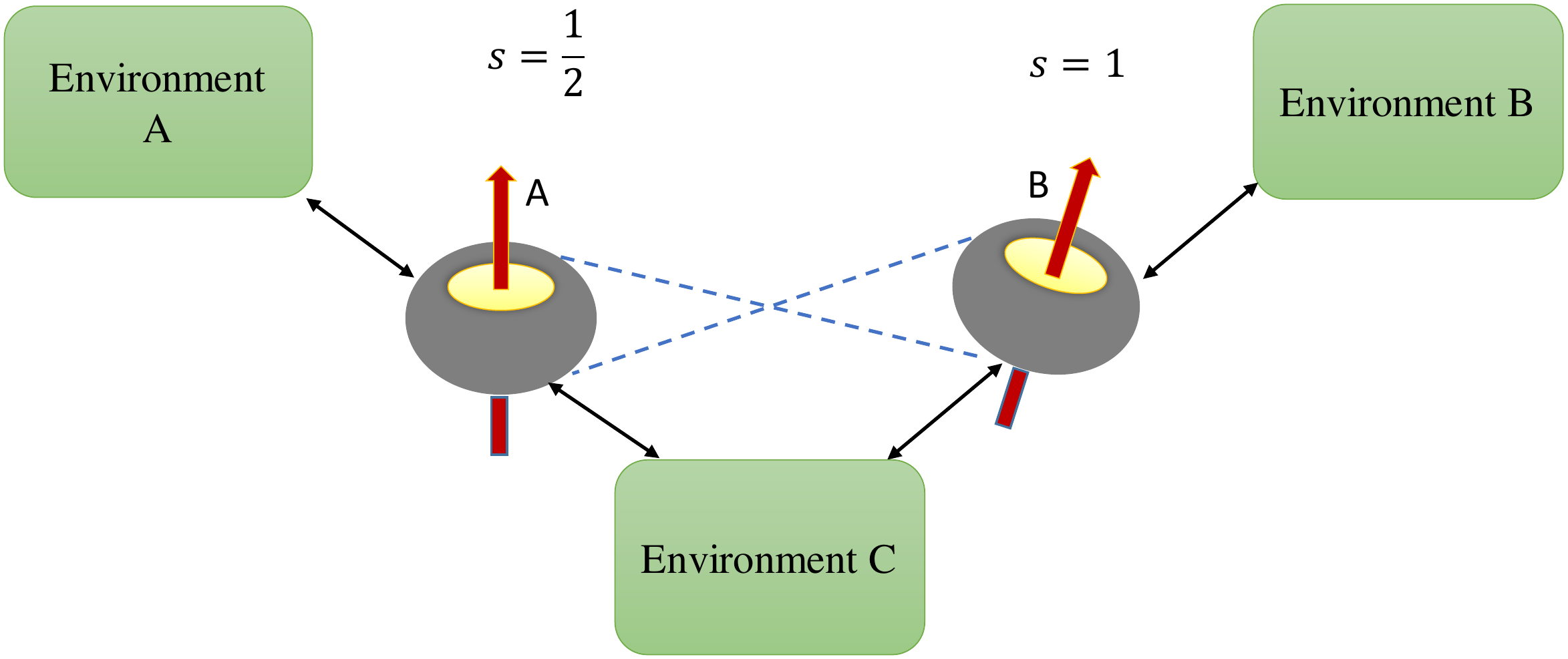}
	\caption{Illustration of the composite qubit($A$)-qutrit($B$) system; Blue dashed lines represent entanglement between the subsystems. The bipartite system can interact either with independent local environments $E_{A}$, $E_{B}$ or with a common environment $E_{C}$.}
	\label{fig:Fig1}
\end{figure}

The composite hybrid qubit($A$)-qutrit($B$) system  consists of  a spin-$\frac{1}{2}$ subsystem (qubit A) and a  spin-1 subsystem (qutrit B). In the following  we study the interaction of this composite system with  local non-Markovian environments  $A$ and $B$ or with common environment $C$ modeling quantum or classical noises. The theoretical schematic of this system is depicted in Fig.~\ref{fig:Fig1}. 

\subsubsection{Coupling to independent squeezed vacuum reservoirs}
Now we investigate the scenario in which each of the subsystems, i.e., the qubit $A$ ${{(s_{A}=\frac{1}{2})}}$ and qutrit $B$ ${{(s_{B}=1)}}$, interacts independently with  its local squeezed vacuum reservoir. For simplicity we assume that  the characteristics of the reservoirs are similar.  Equation (\ref{eq:Eq21}) with the decoherence factor introduced in Eq. (\ref{Squeezdec}) gives the reduced density matrices of the subsystems. Computing them   and applying the method presented in \cite{PhysRevLett.99.160502}, one can obtain the elements of the evolved  density matrix of the composite system as follows \cite{ji2018quantum}
\begin{equation}\label{eq28}
	{\rho_{AB}}_{nm}{(t)}={\rho_{AB}}_{nm}{(0)} \exp{[-{(n_{A}-m_{A})^2}-{(n_{B}-m_{B})^2}]\gamma{(t)}},
\end{equation}
where $n_{A},m_{A}=-s_{A},...,s_{A}$ and $n_{B},m_{B}=-s_{B},...,s_{B}$.

\subsubsection*{Pure initial state}
We take the hybrid qubit-qutrit system initially in the pure state \cite{PhysRevA.102.022221}
\begin{equation}\label{eq:Eq14}
	\vert\psi\rangle=\frac{1}{\sqrt{6}}{\left(e^{i\phi}\vert 00\rangle+\vert01\rangle+\vert02\rangle+\vert10\rangle+\vert11\rangle+\vert12\rangle\right)},
\end{equation}
which leads to a dynamics of the system described by the evolved reduced density matrix $ \rho{\left(t\right)} $
whose elements  are presented in Appendix \ref{Squuezd vacuum reservoirs}.
Then, the HSS is obtained as
\begin{equation}
	HS\!S=\frac{1}{6}\sqrt{2e^{-2\gamma{\left(t\right)}}+e^{-4\gamma{\left(t\right)}}+e^{-8\gamma{\left(t\right)}}+e^{-10\gamma{\left(t\right)}}}.
\end{equation} 
 
 \par
\begin{figure}[H]
	\centering
	\includegraphics[width=0.49\textwidth]{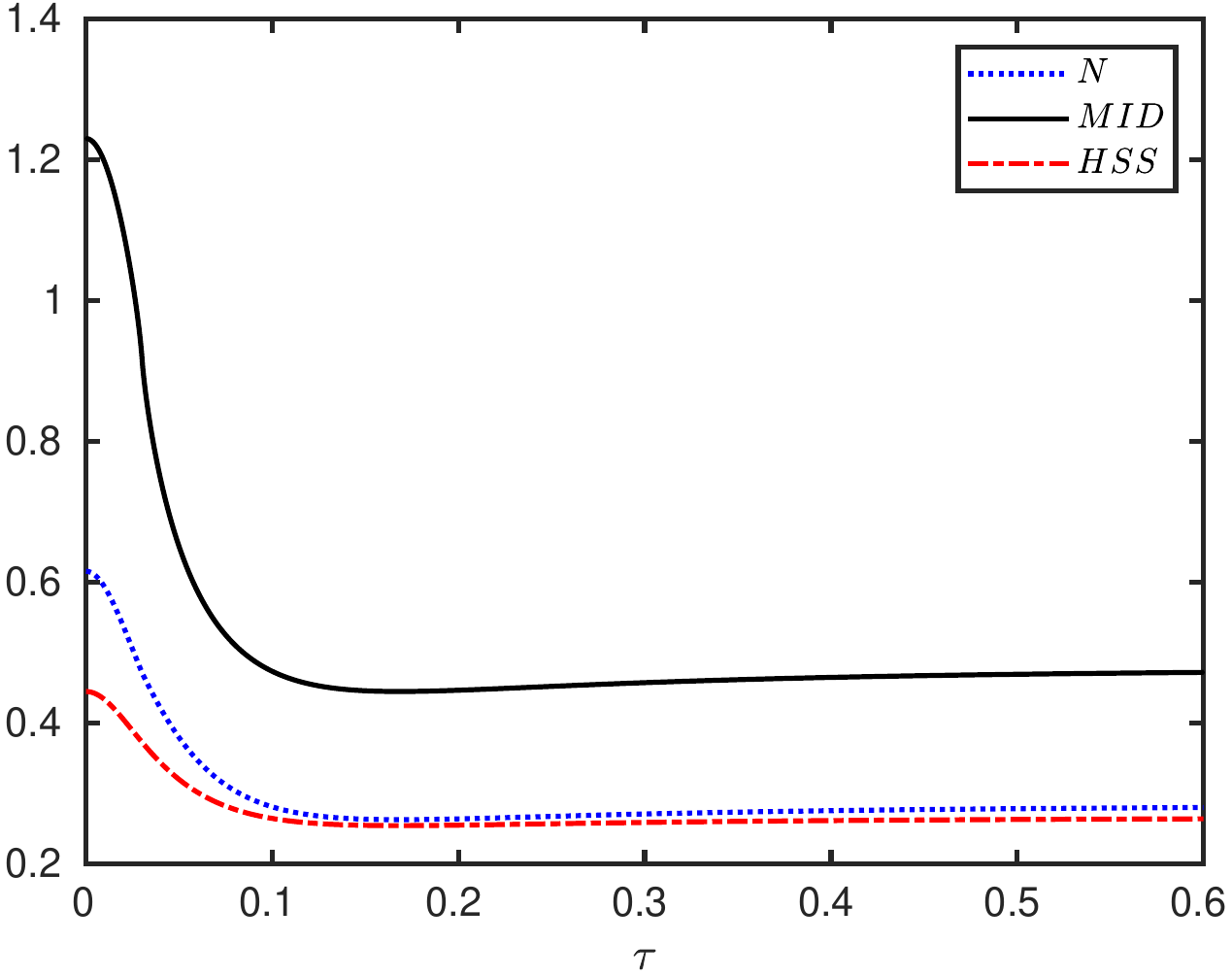}
	\caption{Evolution of the negativity, MID and HSS as a function of dimensionless time $\tau = \omega_0 t$ when each subsystem of the hybrid qubit-qutrit system, starting from the initial pure state, is independently subject to a squeezed vacuum reservoir. The values of the other parameters are $\alpha=0.1$, $\omega_{c}=20\omega_{0}$, $r=0.3$, $ \phi=\pi $ and $s=3$.}
	\label{fig:Fig2}
\end{figure}

The dynamics of negativity, MID and HSS computed by the  evolved state of the system are plotted in Fig.~\ref{fig:Fig2}. We find all measures  initially decrease with time,  then start to increase, and finally exhibit freezing phenomenon \cite{PhysRevLett.114.210401}.
As discussed, the revival of 
the quantum correlation measures can be attributed to the non-Markovian evolution of the system \cite{PhysRevA.85.032318}.
We see  that the behaviors of the HSS, negativity and quantum correlation exhibit an excellent qualitative agreement. Consequently,  the HSS-based witness can precisely capture the
non-Markovian dynamics of  the composite system.

\subsubsection*{Mixed  initial  state}
The non-Markovianity of the system, as faithfully individuated by quantum correlation measures, may in general depend on the initial state. It is thus important to investigate whether the HSS witness, obtained from the initial pure state of Eq.~(\ref{eq:Eq14}) by definition, is capable to identify the non-Markovian character of the system dynamics also when the system starts from a mixed state. We shall study this aspect here and in all the other environmental conditions considered hereafter (see sections below devoted to a mixed initial state).

We consider the one-parameter mixed entangled state as the initial state of the hybrid qubit-qutrit system \cite{KARPAT20114166}
\begin{equation}\label{eq:Eq25}
	\rho_{0}{\left(p\right)}=\frac{p}{2}{\left(\vert 01\rangle\langle01\vert+\vert11\rangle\langle 11\vert\right)}+p\vert\psi^{+}\rangle\langle\psi^{+}\vert+\left(1-2p\right)\vert\psi^{-}\rangle\langle \psi^{-}\vert,
\end{equation}
where
\begin{equation}\label{eq:Eq26}
	\begin{split}
		\vert\psi^{+}\rangle=\frac{1}{\sqrt{2}}{\left(\vert 00\rangle+\vert12\rangle\right)},\\
		\vert\psi^{-}\rangle=\frac{1}{\sqrt{2}}{\left(\vert 02\rangle+\vert10\rangle\right)},
	\end{split}
\end{equation}
in which the entanglement parameter $ p $ varies from $ 0 $ to $ 1 $ such that  $\rho{(p)}$ is  entangled  except  for $p=\frac{1}{3}$. We point out that such a state is taken as the initial state of the system for the dynamics of the quantum correlation quantifiers, namely negativity and MID. 
We find that Eq.~(\ref{eq:Eq25}) leads to the evolved state of the system
\begin{equation}\label{eq:Eq29}
	\rho{(t)} =\left(
\begin{array}{cccccc}\frac{p}{2} & 0 & 0 & 0 & 0 & \frac{p}{2} \mathcal{F} \\
0 & \frac{p}{2} & 0 & 0 & 0 & 0 \\
	0 & 0 & \frac{1-2 p}{2}& \frac{1-2 p}{2} \mathcal{F} & 0 & 0 \\
0 & 0 & \frac{1-2 p}{2} \mathcal{F}  & \frac{1-2 p}{2} & 0 & 0 \\
0 & 0 & 0 & 0 & \frac{p}{2} & 0 \\
\frac{p}{2} \mathcal{F} & 0 & 0 & 0 & 0 & \frac{p}{2} \\
	\end{array}
\right),
\end{equation}
where $\mathcal{F}=e^{-5\gamma{(t)}}$.
Then, the negativity is given by \cite{TCHOFFO20191856}
\begin{equation}\label{eq:Eq30}
	\begin{split}
		\mathcal{N}=&\frac{(p-1)}{2}+\frac{1}{4}{\vert p+(1-p)\mathcal{F}\vert}+\frac{1}{4}{\vert p-(1-p) \mathcal{F}\vert}+\frac{1}{4}{\vert p-(1-2p)\mathcal{F}\vert}+\\&\frac{1}{4}{\vert p+(1-2p)\mathcal{F}\vert}.
	\end{split}
\end{equation}
Moreover, using Eq.~(\ref{MIDformul}) we can compute the MID as
\begin{equation}\label{eq:Eq31}
	\mathcal{M}=\frac{\left(1-p\right)}{2}{\left[ \left(1+\mathcal{F}\right)\log{\left(1+\mathcal{F}\right)}+\left(1-\mathcal{F}\right)
		\log{\left(1-\mathcal{F}\right)}\right]}.
\end{equation}

\begin{figure}[H]
    \centering
	\includegraphics[width=0.49\textwidth]{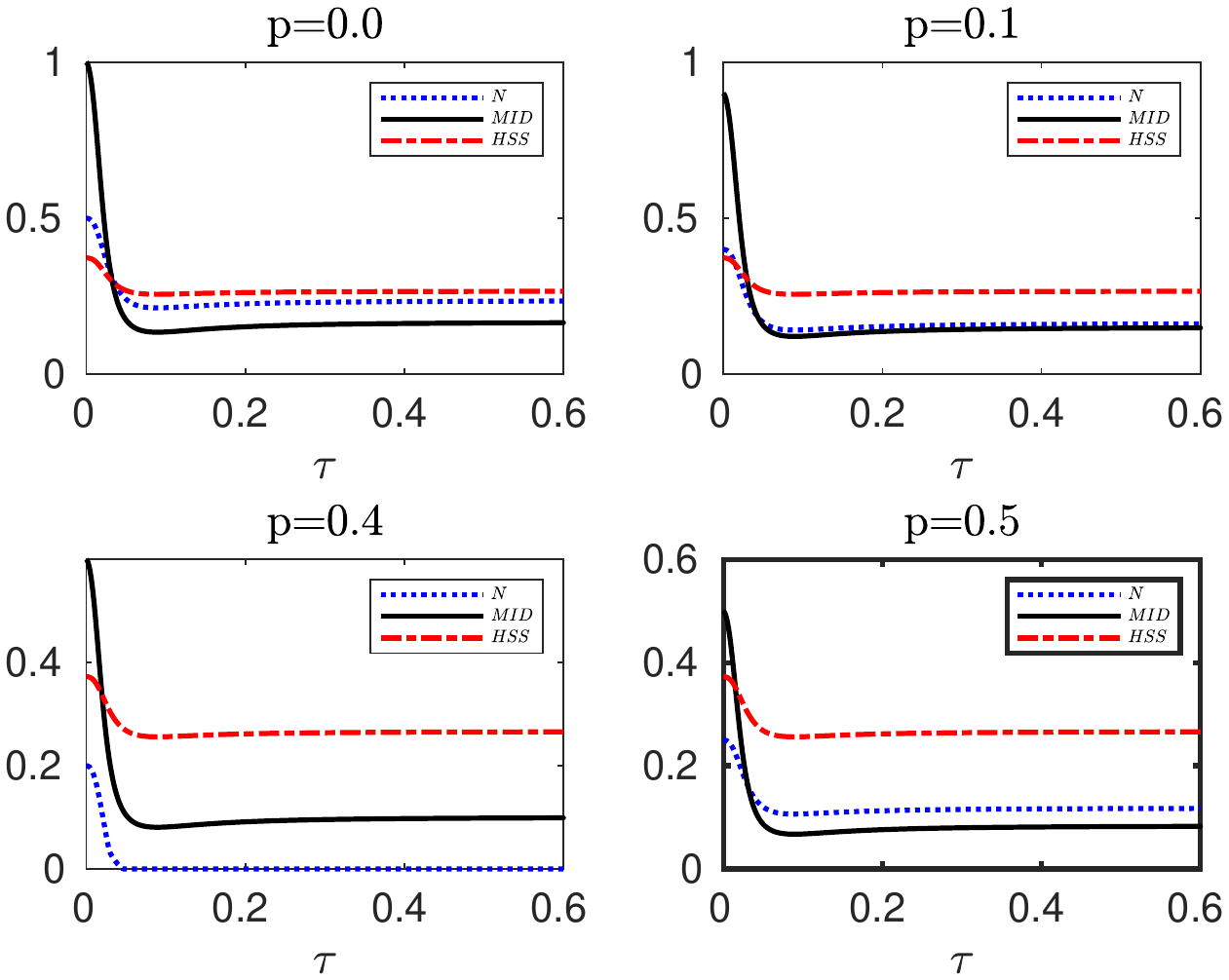}
	\caption{Comparing the evolution of negativity and MID computed for the initial mixed state of the hybrid qubit-qutrit system, when each subsystem is independently coupled to a squeezed vacuum reservoir, with HSS (obtained from the initial pure state) for different values of the entanglement parameter $p$ . In all plots the remaining parameters are $\alpha=0.1$, $s=3$, $\omega_{c}=20\omega_{0}$, $r=0.3$.}
	\label{fig:Fig3}
\end{figure}

In Fig.~\ref{fig:Fig3} we compare the evolution of HSS, obtained from the initial pure state of Eq.~(\ref{eq:Eq14}), with the dynamics of 
negativity and MID, computed for the mixed initial state of Eq.~(\ref{eq:Eq15}), for different values of $p$. 
The dynamics of the HSS is again in perfect agreement with that observed for the entanglement and quantum correlations as quantified by the negativity and  MID, respectively. Therefore, the HSS-based witness, computed versus the phase parameter encoded into an initial pure state of the system, can efficiently detect the non-Markovian dynamics even in the case when the initial state of our high-dimensional system is not pure.
 It should be noted that in the presence of sudden death of entanglement, which occurs   for some values of the entanglement parameter (for example, for $p=0.4$), only  the  HSS and  MID show the same dynamics. Hence, the negativity cannot be used as a faithful witness of non-Markovianity when it exhibits the sudden death phenomenon.

In  the case of initially entangled noninteracting qubits
 in independent non-Markovian quantum environments, entanglement or quantum correlation  revivals can be explained in terms of transfer
 of correlations back and forth from the composite system
 to the various parts of the total system. This is due to
 the back-action via the environment on the system, which
 creates correlations between qubits and environments and
 between the environments themselves. Accordingly, in this case the non-Markovianity is defined as backflow of information  from the environment(s) to the
 system(s).
 
\subsubsection{Coupling to classical environments}\label{classicalenv}
Here we assume that the hybrid qubit-qutrit system is affected by a  classical environment implemented by the random telegraph noise (RTN) with a Lorentzian spectrum. It is a famous class of non-Gaussian noises  used to generate the low-frequency $\frac{1}{f^{\alpha}}$ noise both theoretically and experimentally. It is also  responsible for coherent
dynamics in quantum solid-state nanodevices \cite{PhysRevLett96097009,PhysRevB79125317,PhysRevA87052328}. Physically,
the RTN may result  from one of the following scenarios: (i) charges flipping between two locations in space (charge noise); (ii) electrons trapping in shallow
subgap formed at the boundary between a superconductor and an
insulator (noise of critical current); and (iii) spin diffusion on a superconductor surface generated by the exchange mediated by the
conduction electrons (flux noise) \cite{PhysRevLett.100.227005,PhysRevLett.97.167001}. The Hamiltonian of the qubit-qutrit system under the RTN is given by
\begin{equation}\label{eq:Eq32}
\begin{split}
    \mathcal{H}{\left(t\right)}=&\mathcal{H}_{0}+\mathcal{H}_{I}\\
	\mathcal{H}_{0}=&\sum_{k=A,B}\epsilon_{k} {S^{Z}_{k}},\mathcal{H}_{I}= \sum_{k=A,B} \left[{J_{k}L_{k}{(t)}+J_{c}C{(t)}}\right]{S^{k}_{z}},
\end{split}
\end{equation}
where $\epsilon_{k}$ denote the energy of an isolated qubit (qutrit), $S_{z}^{A}=\sigma_{z}$ and $S_{z}^{B}$ represent the spin operators of, respectively, the qubit and the
qutrit in the $z$-direction. Moreover, $J_{k}$ and $J_{c}$ represent the coupling strengths of each marginal system to
the local and non-local RTN, such that we consider two types of the system-environment interactions, namely
\begin{itemize}
\item[1)] local or independent environments $(ie)$:~$J_{k}=\nu\neq0$ and $J_{c}=0$;
\item[2)] non-local or common environments $(ce)$:~$J_{k}=0$ and $J_{c}=\nu\neq0$.
\end{itemize}
Furthermore, $L_{k}{(t)}$ and $C(t)$ denote the
random variables used to introduce the stochastic processes. They 
are used  to describe the different conditions under which
the subsystems undergo the decoherence due to the environment.
Here, they
represent  classical random fluctuating fields
such as  bistable fluctuators flipping between two fixed values $\pm m$
at rates $\gamma_{k}$ and $\gamma$, respectively. For simplicity, we assume that $\gamma_{k}=\gamma$. For the \textit{autocorrelation function} of the random variable $\eta{(t)} ={\{L_{k}{(t)}; C(t) \}}$ we have
$\langle \delta \eta{(t)}\delta \eta{(t')}\rangle= \exp \left[-2 \gamma \vert t-t' \vert \right]$ with a Lorentzian power spectrum 
$S{(\omega)} = \frac{4\gamma}{\omega^{2}+\gamma^{2}}$. Defining the parameter $q = \frac{\gamma}{\nu}$, we can identify two
regimes for the dynamics of quantum correlations:
 the Markovian regime ($q\gg 1$: fast RTN), and the non-Markovian regime ($q\ll 1$: slow RTN).
The time-evolving state of the system under the influence of the RTN is given by
\begin{equation}
	\rho{\left(\{\eta\},t\right)}=U{\left(\{\eta\},t\right)}\rho{\left(0\right)} U^{\dagger}{\left(\{\eta\},t\right)}.
\end{equation}
in which the time-evolution operator $U{\left(\{\eta\},t\right)}$ called the stochastic unitary operator in the interaction
picture is given by
\begin{equation}\label{eq:Eq34}
	U{\left(\{\eta\},t\right)}=\exp{\left[-i\int_{0}^{t}\mathcal{H}_{I}{\left(t'\right)}dt'\right]}.
\end{equation}
where $\eta{(t)} ={\{L_{k}{(t)}; C(t) \}}$  stands for the different realizations of the stochastic process.
Because  $ U{\left(\{\eta\},t\right)} $ depends on the noise, we should perform the ensemble average over the
noise fields to obtain the reduced density matrix of the open system,  i.e., 
\begin{equation}
	\rho_{ie(ce)}=\langle \rho{\left(\{\eta\},t\right)} \rangle_{\eta{(t)}}.
\end{equation}
The evolved state of the system in the presence of independent environments (ie) and collective environments (ce) is obtained as
\begin{equation}\label{eq:Eq38}
	\begin{split}
		\rho_{ie}{(t)}=\langle{\langle \rho{(\theta_{A}{(t)},\theta_{B}{(t)},t)}\rangle_{\theta_{A}}}\rangle_{\theta_{B}}\\
		\rho_{ce}{(t)}=\langle \rho{(\theta{(t)},t)}\rangle_{\theta},
	\end{split}
\end{equation}
where $\theta_{k}{(t)}=\nu \int_{0}^{t}L_{k}{(t')}dt'$  ($ k=A,B $) and $\theta{(t)}=\nu \int_{0}^{t}C{(t')}dt$.
Calculation  of the above terms  requires the computation  of averaged terms of the
type $\langle e^{\pm in\theta}\rangle$ $\left(n\in N \right)$  given by
 \cite{2017QuIP16191T}
\begin{equation}
	\begin{split}
		\langle e^{in\theta}\rangle=D_{n}{(\tau)}=\langle \cos{\left(n\theta\right)}\rangle\pm i\langle\sin{\left(n\theta\right)}\rangle,\\
		\langle\sin{\left(n\theta\right)}\rangle=0,
	\end{split}
\end{equation}
\[
\langle \cos{\left(n\theta\right)}\rangle=
\begin{cases}
	e^{-q\tau}{\left[\cosh{\left(\xi_{qn}\tau\right)}+\frac{q}{\xi_{qn}}\sinh{\left(\xi_{qn}\tau\right)}\right]},& \text{ q$>$n}\\
	e^{-q\tau}{\left[\cos{\left(\xi_{nq}\tau\right)}+\frac{q}{\xi_{nq}}\sin{\left(\xi_{nq}\tau\right)}\right]},              & \text{q $<$n}
\end{cases}
\]
where $\xi_{ab}=\sqrt{a^2-b^2}$ $((a,b) = n,q)$, and $\tau=\nu t$ denotes the scaled (dimensionless) time \cite{tchoffo2019frozen}.

\subsubsection*{Pure initial  state in the presence of independent classical environments}
Here we assume that each of  the qubit and  the qutrit  locally interacts with its local RTN while the composite system starts with the pure initial state (\ref{eq:Eq14}).
For this case the elements of evolved density matrix are given in Appendix \ref{Pure qubit-qutrti_RTN}.
Then the HSS is obtained as
\begin{equation}
	\begin{split}
		HS\!S=\frac{1}{6}{\sqrt{D_{1}^{2}{\left(\tau\right)}+2D_{2}^{2}{\left(\tau\right)}+D_{2}^{2}{\left(\tau\right)}D_{1}^{2}{\left(\tau\right)}+D_{2}^{4}{\left(\tau\right)}}}.
	\end{split}
\end{equation}

\begin{figure}[t!]
	\centering
	\includegraphics[width=0.49\textwidth]{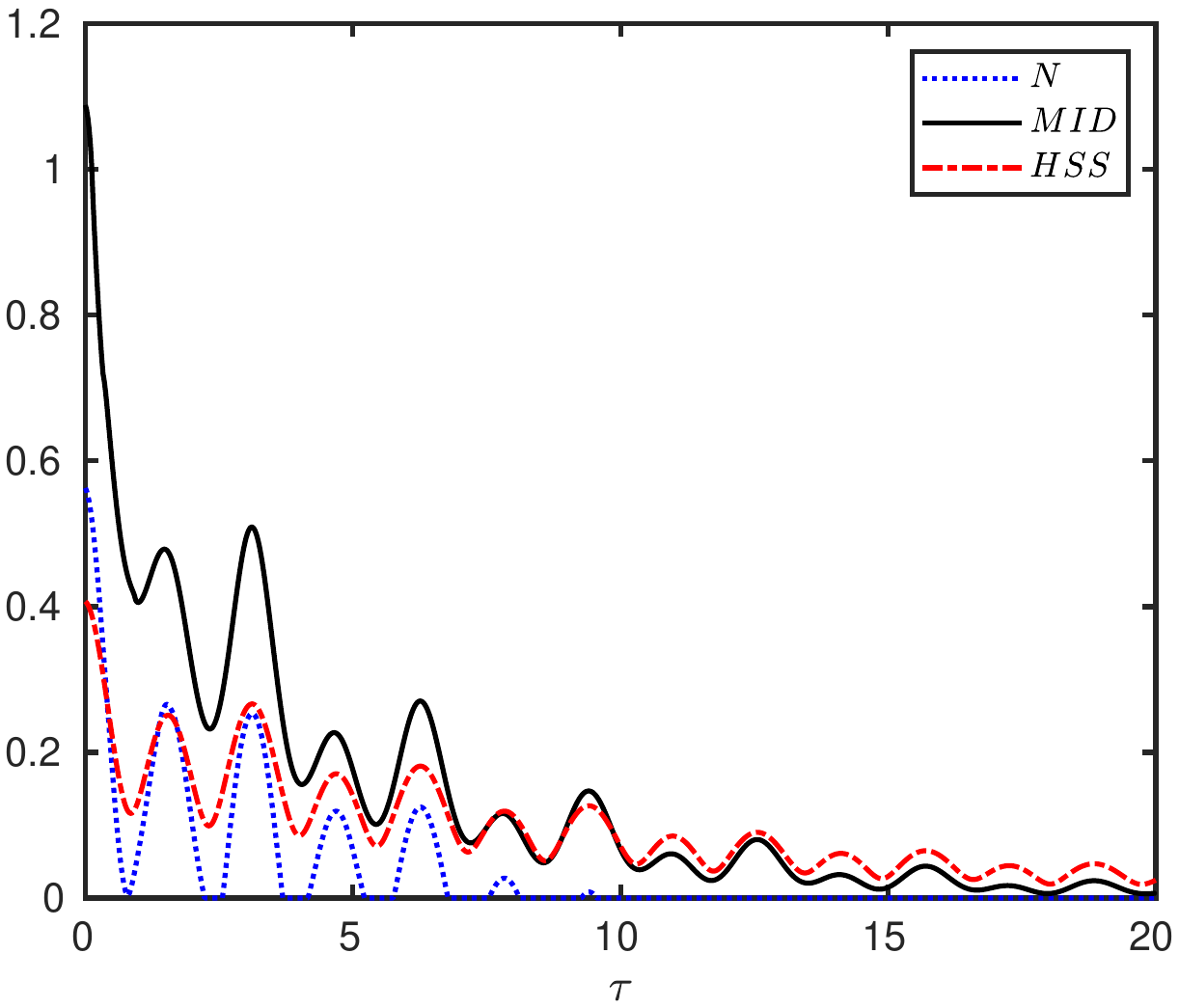}
	\caption{Evolution of negativity, MID and HSS as a function of dimensionless time $\tau = \nu t$ when each subsystem of the hybrid qubit-qutrit system, starting from the initial pure state, is independently subject to a random telegraph noise in non-Moravian regime $q=0.1$.}
	\label{Fig4}
\end{figure}

In Fig.~\ref{Fig4} we illustrate the time behaviors of the negativity, MID and HSS  in the non-Markovian regime as a function of the dimensionless time. It is clear  that when the entanglement sudden death occurs, the HSS and MID synchronously oscillate with time as they are suppressed to the minimum value and then rise. Moreover, at the first revival of the measures, the minimum  point of the  HSS exactly coincide with that of the negativity. After that moment we see that  maximum (minimum) points of the HSS are in complete coincidence with maximum (minimum) points of the negativity as well as the MID. This perfect qualitative agreement is an evidence that the HSS-based witness can precisely detect the non-Markovianity in the presence of classical noises.

\subsubsection*{Mixed initial  state in the presence of independent classical environments}
Now we  compare the dynamics of the HSS, obtained from the initial pure state of Eq.~(\ref{eq:Eq14}) by definition, with the evolution of the negativity and quantum correlation computed for the initial mixed state of Eq.~(\ref{eq:Eq25}). The evolved density matrix,  the corresponding negativity and quantum correlation are obtained from, respectively, Eqs. (\ref{eq:Eq29}-\ref{eq:Eq31}) replacing $\mathcal{F}$ with ${D_{2}{(\tau)}}^{2}$.

\begin{figure}[H]
	\centering
	\includegraphics[width=0.49\textwidth]{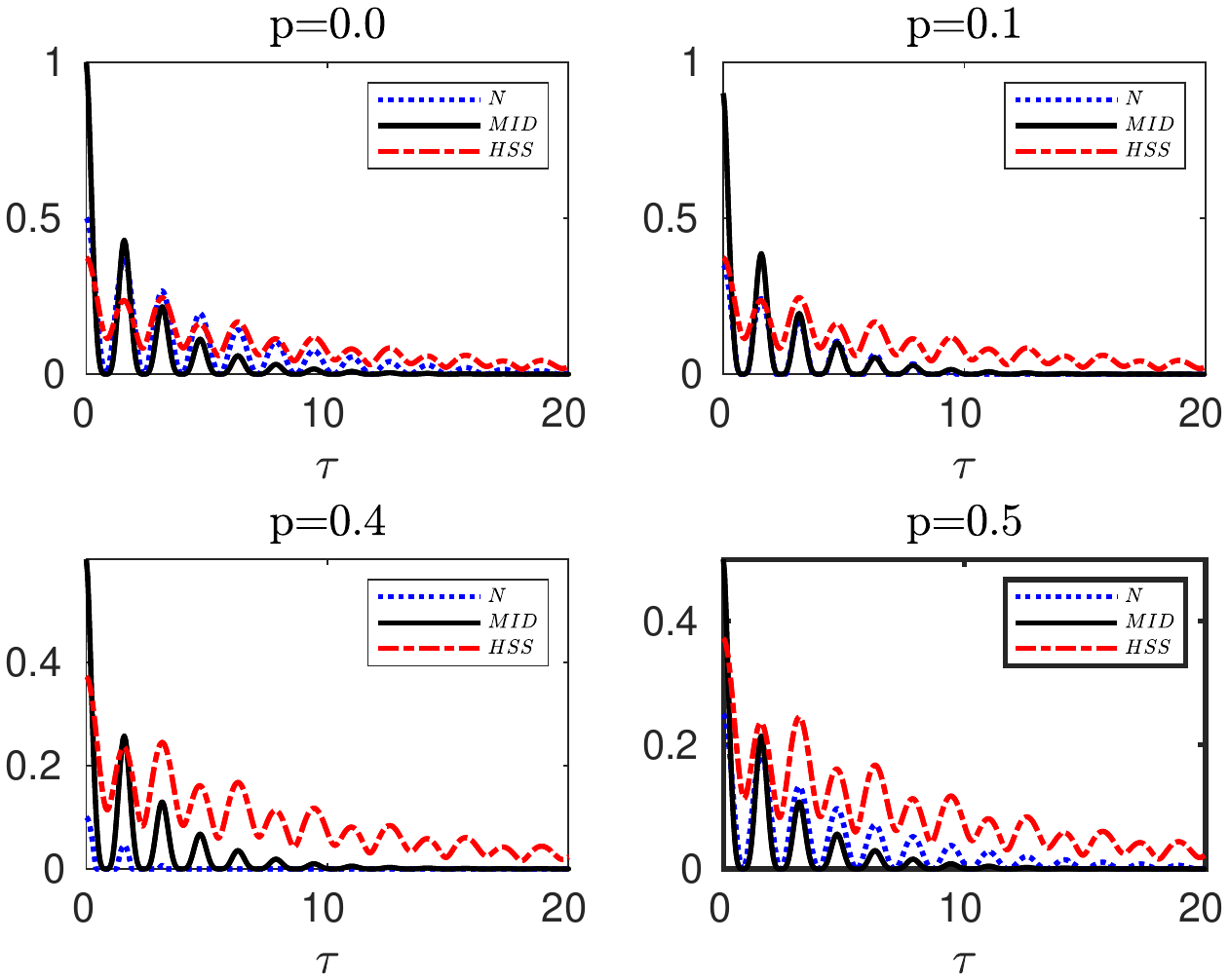}
	\caption{Comparing the evolution of negativity and MID  computed for the initial mixed state of the hybrid qubit-qutrit system, when each subsystem is independently coupled to a random telegraph noise, with HSS (obtained from the initial pure state) for different values of the entanglement parameter $p$ in the non-Markovian regime: $ q = 0.1 $.}
	\label{fig:Fig5}
\end{figure}

Figure~\ref{fig:Fig5} exhibits this comparison for  different values of the entanglement parameter $p$. Not considering the periods when the sudden death of the entanglement occurs, we observe that the maximum and minimum points of the measures are very close to each other  and small deviations originate from the fact that the initial state, used for computation of the HSS-based measure, should be optimized over all possible
parametrizations. Therefore, the HSS-based measure remains as  a valid non-Markovianity
identifier in the presence of the classical noises.

%It is worth to mention that for entanglement parameter of $p=0$ and $p=\frac{1}{2}$, HSS computed for initial pure state is in more qualitative agreement with negativity obtained from initial mixed state than quantum correlation, however for other values of $p$ like $p=0.1$ and $p=0.4$ make the initial state little entangled with an early stage disentanglement, negativity goes to zero quickly for ($p=0.4$), then in this case HSS is in qualative agreement with quantum correlation. So, it's reasonable that HSS in this case agrees with the dynamics of more general quantum correlations (MID, or discord-like correlations). It remains a signature of non-Markovianity due to the nonmonotonicit oscillation of HSS.

\subsubsection*{Mixed initial  state  in the presence of a common classical environment}
\begin{figure}[t!]
	\centering
	\includegraphics[width=0.49\textwidth]{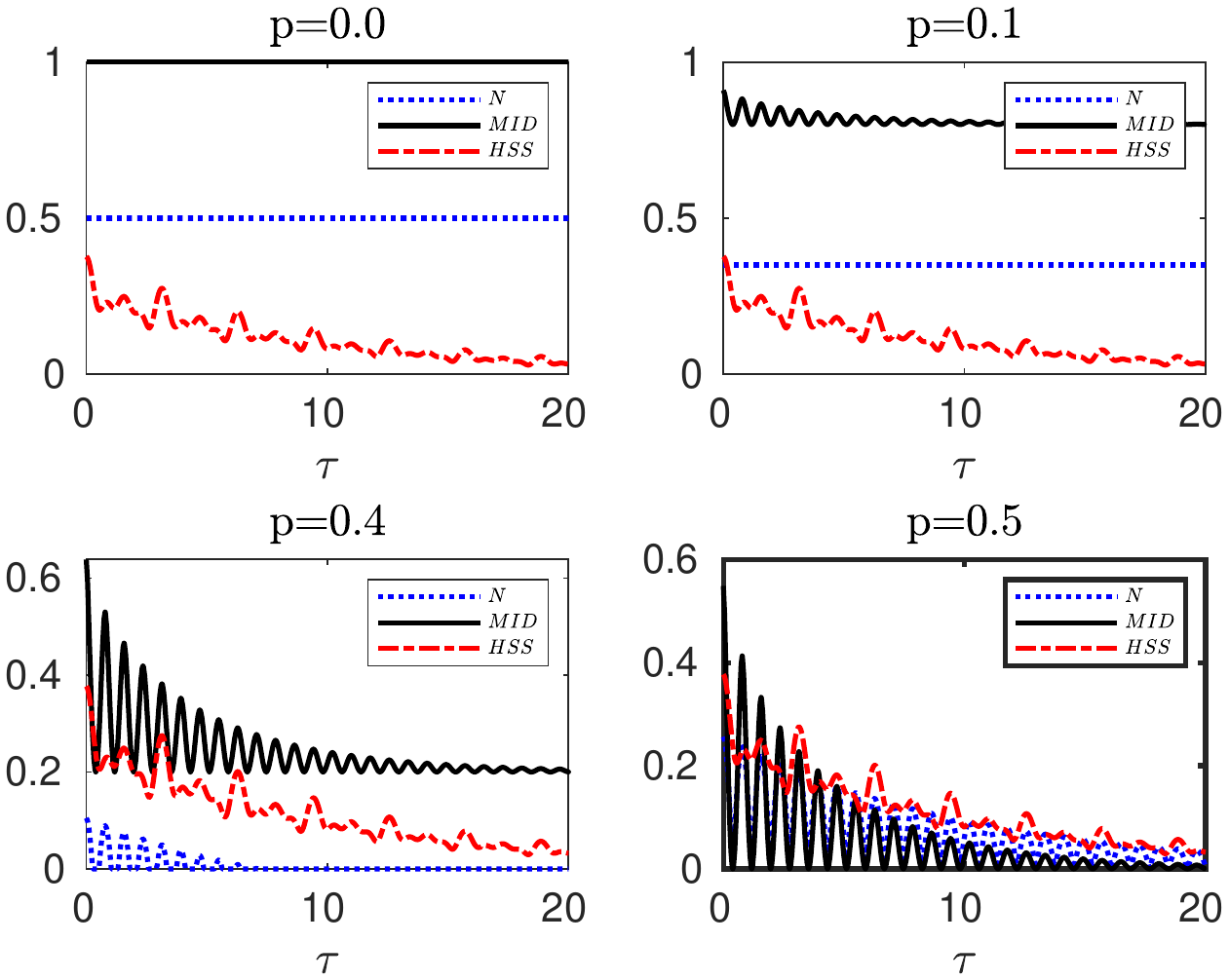}
	\caption{Comparing the evolution of negativity and MID  computed for the initial mixed state of the hybrid qubit-qutrit system, when its subsystems are subject to a common RTN source, with HSS (obtained from the initial pure state) for different values of the entanglement parameter $p$ in the non-Markovian regime: $ q = 0.1 $. }
	\label{fig:Fig6}
\end{figure}

Let us now compare the dynamics of the HSS, obtained as usual from the initial pure state of (\ref{eq:Eq14}) by definition, with the evolution of  the negativity and quantum correlation computed for the initial mixed state of Eq.~(\ref{eq:Eq25}),
when both the qubit and the qutrit are embedded into a common RTN source in the non-Markovian regime. The elements of the evolved dynamical density matrix are given in Appendix \ref{pure_qubit_qutrit_common_rtn}. Then, one can easily determine the HSS as 
\begin{equation}
    HSS=\frac{1}{6}{\sqrt{{D_{1}{(\tau)}}^{2}+2{D_{2}{(\tau)}}^{2}+{D_{3}{(\tau)}}^{2}+{D_{4}{(\tau)}}^{2}}}.
\end{equation}
Moreover, the evolved density matrix  of  the  hybrid qubit-qutrit system for the  initial mixed state of Eq.~(\ref{eq:Eq25}) is obtained as
\begin{equation}\label{eq;Eq41}
	\rho{(t)} =\left(
	\begin{array}{cccccc}\frac{p}{2} & 0 & 0 & 0 & 0 & \frac{p}{2} \mathcal{F}e^{i\phi} \\
		0 & \frac{p}{2} & 0 & 0 & 0 & 0 \\
		0 & 0 & \frac{1-2 p}{2}& \frac{1-2 p}{2}  & 0 & 0 \\
		0 & 0 & \frac{1-2 p}{2}  & \frac{1-2 p}{2} & 0 & 0 \\
		0 & 0 & 0 & 0 & \frac{p}{2} & 0 \\
		\frac{p}{2} \mathcal{F}e^{-i\phi} & 0 & 0 & 0 & 0 & \frac{p}{2} \\
	\end{array}
	\right),
\end{equation} 
where $\mathcal{F}=D_{4}{(\tau)}$. 
As a consequence, we find that the negativity and MID are, respectively, 
\begin{equation}
	\begin{split}
		\mathcal{N}=
		&\frac{1}{4}{\left[
		(p-1)+\vert{3p-1}\vert+\vert{(1-2p)-p\mathcal{F}}\vert+\vert{(1-2p)+p\mathcal{F}}\vert\right]
		},
	\end{split}
\end{equation}
\begin{equation}
	\begin{split}
		\mathcal{M}=&\left(1-2p\right)+
		\frac{p}{2}\left(1+\mathcal{F}\right)\log{\left(1+\mathcal{F}\right)}+\frac{p}{2}\left(1-\mathcal{F}\right)
		\log{\left(1-\mathcal{F}\right)}.
	\end{split}
\end{equation}
A common environment induces a mutual interaction between  subsystems which may lead to the preservation of correlations in some cases \cite{PhysRevLett.91.070402,PhysRevLett.89.277901,10.1142/S0217979213450537} (see  Fig. \ref{fig:Fig6} demonstrating this feature of common environments causing the MID to fail in  detecting the non-Markovianity). Except for these situations, we see  that the maximum  (minimum) points of the HSS computed for the initial pure state are very close to those of the MID calculated for the initial mixed state. Hence, the 
 HSS-based measure can be used as a faithful witness of non-Markovianity when the subsystems are affected by a common classical noise.

It should be noted that the classical
environments cannot store any quantum correlations on their
own, and hence they do not become entangled with their respective
quantum systems. Accordingly, common interpretation of non-Markovianity in accordance with inflow (outflow) of information to (from) the system may be problematic in the presence of the RTN and other similar classical noises \cite{franco2012revival}. In other words, although a general classical environment may hold information about the
quantum systems, it is somewhat misleading to talk about information flow from the system(s) to the environment(s) or information backflow from the environment(s) to the system(s).
The better interpretation is to say that the quantum system
has a recording memory of the events affecting its dynamics. 
When the quantum memory starts remembering, the information about the past events becomes accessible, leading to revival of the quantum correlations and hence to the appearance of quantum non-Markovianity \cite{jahromi2015precision}.

\subsubsection{Composite classical-quantum environments}
Here we investigate a hybrid system formed by a qubit  subjected to a random telegraph noise and a qutrit independently subjected to a squeezed vacuum reservoir. The Hamiltonian of such a system can be written as
\begin{equation}\label{eq:Eq96}
	\mathcal{H}=\mathcal{H}_\mathrm{qb}(t)\otimes {\mathcal{I}}_\mathrm{qt}+{\mathcal{I}}_\mathrm{qb} \otimes\mathcal{H}_\mathrm{qt}(t).
\end{equation}
where ${\mathcal{I}}_\mathrm{qb(qt)}$ denotes the identity operator acting on the subspace of the qubit (qutrit). Moreover,  the Hamiltonians of the local interaction of the qubit and qutrit,  $\mathcal{H}_{qb}(t)$ and $\mathcal{H}_{qt}(t)$, as well as their corresponding evolution operators, $ {\mathcal{U}_{qb}{\left(\theta,t\right)}} $ and $ {\mathcal{U}_{qt}{\left(\theta,t\right)}}  $ can be extracted from  Secs. \ref{classicalenv} and \ref{Example}. In addition, one can consider the unitary evolution operator of the system as
$
    \mathcal{U}={\mathcal{U}_{qb}{\left(\theta,t\right)}}\otimes{\mathcal{U}_{qt}{\left(t\right)}}.
$
%where ${\mathcal{U}_{qt (qb)}{\left(\theta,t\right)}}$ is given in the form Eq. $\left(\ref{eq:Eq15}\right) (\left(\ref{eq:Eq34}\right))$.
Then, the evolved density matrix  of the this system can then be  obtained  by averaging the unitary
evolved density matrix  over the stochastic process induced by the RTN.
%\begin{equation*}
 %   \rho{\left(t\right)}=\langle{\langle { \mathcal{U}{\left(\theta,t\right)} \rho_{0}\mathcal{U}^{\dagger}{\left(\theta,t\right)}}\rangle}_{\theta}\rangle_{B},
%\end{equation*}
%then the elements of density matrix can be considered as 
%\begin{equation}\label{eq:Eq44}
%\begin{split}
 %   \rho_{nm}{\left(t\right)}=\alpha_{nm}\langle{ \mathcal{U}_{qb}{\left(\theta,t\right)} \ket{n_{A}}\bra{m_{A}}{\mathcal{U}_{qb}}^{\dagger}{\left(\theta,t\right)}}\rangle_{\theta}\\\times
    % \exp{[-(n_{B}-m_{B})^2\gamma(t)]},
%\end{split}
%\end{equation}
%where $\alpha_{nm}$ is the amplitude of $\rho_{nm}\left(0\right)$ and $n_{A}$, $m_{A}$,  $n_{B}$, $m_{B}$ are computational basis of qubit and qutrit.

\subsubsection*{Pure initial state}
\begin{figure}[t!]
	\centering
	\includegraphics[width=0.49\textwidth]{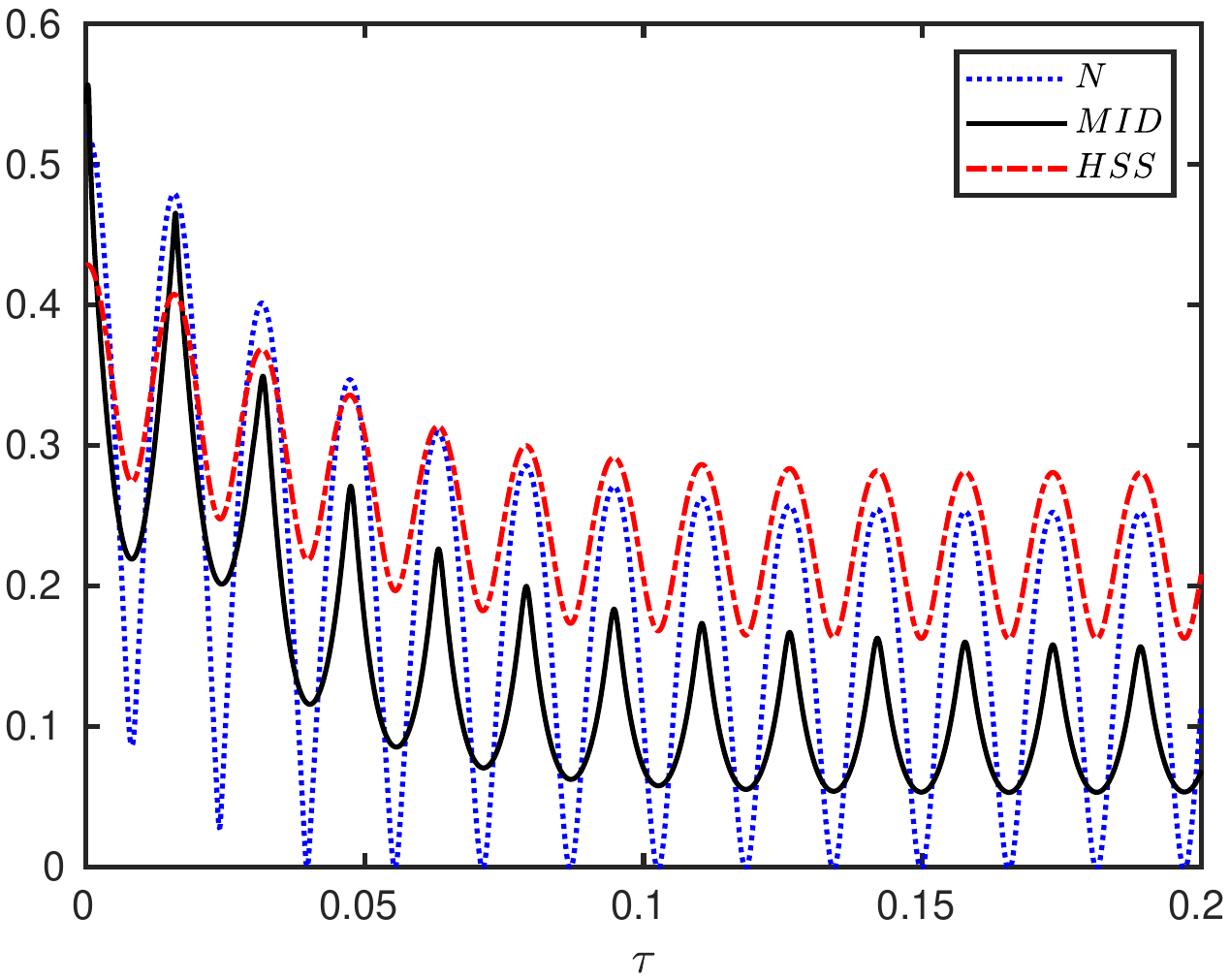}
	\caption{Evolution of negativity, MID and HSS as a function of dimensionless time $\tau $ when  the subsystems of the hybrid qubit-qutrit system,  starting from the initial pure state, are independently subject to composite classical-quantum environments. The values of the other parameters are given by  $\alpha=0.1$, $\omega_{c}=20\omega_{0}$, $r=0.3$, and $\nu=100$.}
	\label{fig:Fig7}
\end{figure}

The elements of  the evolved density matrix when starting from the pure state of Eq.~(\ref{eq:Eq14}) are given in Appendix \ref{Pure hybrid qubit-qutrit subject composite environments}, leading to the following expression for the  HSS:
\begin{equation}
	HSS=\frac{1}{6}\sqrt{\left(e^{-2\gamma{\left(t\right)}}+e^{-8\gamma{\left(t\right)}}\right){\left(1+D_{2}{\left(\tau\right)}^{2}\right)}+D_{2}{\left(\tau\right)}^{2}}.
\end{equation}
The time behaviors of  negativity, MID and HSS are shown in Fig.~\ref{fig:Fig7} illustrating that  all measures  exhibit simultaneous oscillations with time such that their maximum and minimum points exactly coincide. This excellent agreement confirms the faithfulness of the HSS-based measure to detect memory effects.

\subsubsection*{Mixed initial state}
\begin{figure}[H]
		\centering
		\includegraphics[width=0.49\textwidth]{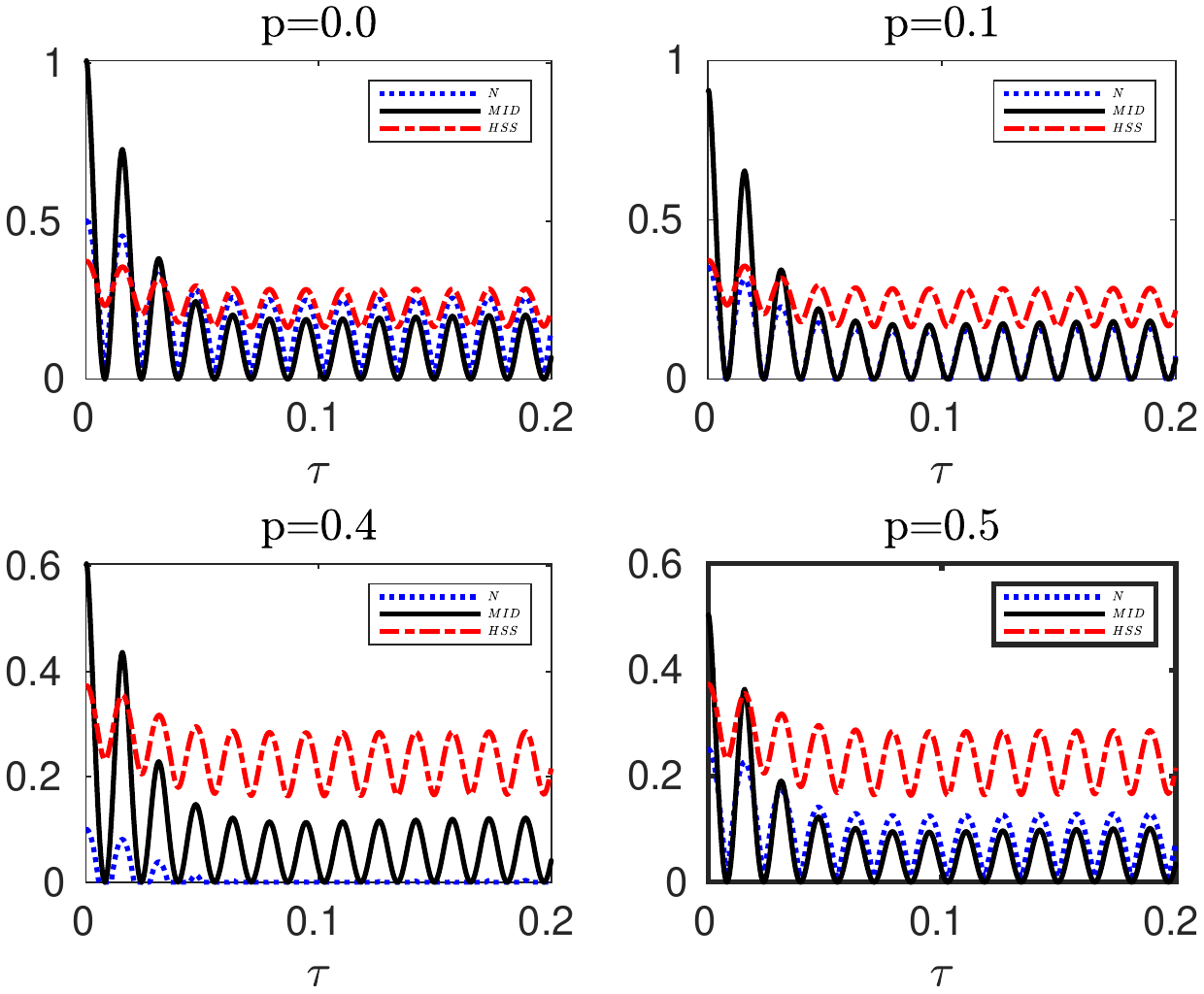}
		\caption{Comparing the evolution of the negativity and MID,  computed for the initial mixed state of the hybrid qubit-qutrit system, when  the subsystems  are independently subject to composite classical-quantum environments,
		 with the HSS obtained from the initial pure state for different values of the entanglement parameter $p$ in the
			non-Markovian regime: $ q = 0.1 $. The values of the other parameters are given by$\alpha =0.1$, $s=3$,$\omega_{c}=20\omega_{0}$, $p=0$ and $v=100$.}
		\label{fig:Fig8}
	\end{figure}

Using Eq.~(\ref{eq:Eq25}) as the initial state and computing the evolved state of the system (See Appendix \ref{APPB4}), we find that the  the negativity and MID, respectively,  are in the  form of Eqs.~(\ref{eq:Eq30}) and (\ref{eq:Eq31}) with $\mathcal{F}=D_{2}(\tau)e^{-4\gamma(t)}$.
In Fig.~\ref{fig:Fig8} the dynamics of negativity and  MID, obtained for the initial mixed state, has been compared with that of the HSS (computed for the initial pure state) in the non-Markovian regime.

The analyses are similar to ones in other discussed scenarios, showing that the HSS-based witness may be a proper non-Markovianity identifier even if the initial state of high dimensional systems is not pure.

%%%%%%%%%%%%%%%%%%%%%%%%%%%%%%%%%%%%%%%%%%
\section{Conclusions}\label{IV}

Recently, the HSS-based witness, a quantifier of quantum statistical speed which has the advantage of avoiding the diagonalization of the evolved density matrix, has been introduced as a trustful witness of non-Markovianity in low-dimensional systems \cite{PhysRevA.102.022221}. 
In this work we have generalized this result showing that the proposed witness is a bona-fide identifier of non-Markovianity for high-dimensional and multipartite open quantum systems. 
This result stems from the observation that the HSS-based witness is in perfect agreement with established non-Markovianity identifiers based on the dynamical breakdown of monotonicity for quantum information resources, such as negativity and measurement-induced disturbance. 
We have found that, despite the common interpretation of non-Markovianity in terms of backflow of information from the environment to the system may be problematic  \cite{banacki2020information}, the HSS-based witness is capable to detect memory effects of the evolved quantum system.

% In fact, our detailed analysis, obtained through several paradigmatic open quantum systems including one qudit (spin-S systems) undergoing squeezed vacuum reservoir, two type pure and mixed hybrid qubit-qutrit coupled to either independent or collective quantum, classical random telegraph noise and composite environment. Therefore positive value of $\frac{\partial{HSS}}{\partial{t}}$ as consequence of nonmonotonic behavior of HSS in qualitative excellent agreement with nonmonotonic behaviors of quantum negativity measure and quantum correlation measure, provide a strong evidence for sensitivity of our HSS-based witness detecting the memory effects of high dimensional open quantum systems.

In order to construct a non-Markovianity measure on the basis of a geometric distance between two quantum states, one of desirable properties is that the distance is contractive, i.e., nonincreasing under any completely positive trace preserving (CPTP) map. It has been shown that  the HSS is contractive under CPTP maps in low-dimensional Hermitian systems \cite{PhysRevA.102.022221}. Checking all of the dynamical cases presented here, we have found that the contractivity of the HSS holds not only in low dimensional systems but also in high-dimensional ones. Our results also motivate further studies about HSS applications in detecting non-Markovianity in continuous variable systems.

\appendix

\section{Pure hybrid qubit-qutrit evolved density matrix}\label{Pure hybrid qubit-qutrit 1}
This appendix presents the elements of the evolved density matrix of hybrid qubit-qutrit system, starting from the initial pure state of Eq.~(\ref{eq:Eq14}), in the presence of quantum and classical noises. This evolved state is required for the assessment of non-Markovianity via the HSS-based witness.

\subsection{Squeezed vacuum reservoirs}\label{Squuezd vacuum reservoirs}
The elements of the evolved  density matrix, when each subsystem of the hybrid qubit-qutrit system is independently subject to a squeezed vacuum reservoir, in the computational basis ${\ket{00},\ket{01},\ket{02},\ket{10},\ket{11},\ket{12}}$ are given by
\begin{gather}
    \rho_{11}{\left(t\right)}=\rho_{22}{\left(t\right)}=\rho_{33}{\left(t\right)}=\rho_{44}{\left(t\right)}=\rho_{55}{\left(t\right)}=\rho_{66}{\left(t\right)}=\frac{1}{6}\nonumber\\
    \rho_{12}{\left(t\right)}=\rho_{14}{\left(t\right)}=\rho_{21}^{*}{\left(t\right)}=\rho_{41}^{*}{\left(t\right)}=\frac{1}{6}e^{i\phi}e^{-\gamma{\left(t\right)}} \nonumber\\ 
    \rho_{13}{\left(t\right)}=\rho_{31}^{*}{\left(t\right)}=\frac{1}{6}e^{i\phi}e^{-4\gamma{\left(t\right)}}\nonumber\\
    \rho_{15}{\left(t\right)}=\rho_{51}^{*}{\left(t\right)}=\frac{1}{6}e^{-2\gamma{\left(t\right)}}\nonumber\\
    \rho_{16}{\left(t\right)}=\rho_{61}^{*}{\left(t\right)}=e^{i\phi}e^{-5\gamma{\left(t\right)}}\nonumber\\
    \rho_{23}{\left(t\right)}=\rho_{25}{\left(t\right)}=\rho_{32}{\left(t\right)}=\rho_{36}{\left(t\right)}=\rho_{45}{\left(t\right)}=\rho_{52}{\left(t\right)}=\rho_{54}{\left(t\right)}=\rho_{56}{\left(t\right)}\\
    =\rho_{63}{\left(t\right)}=\rho_{65}{\left(t\right)}=\frac{1}{6}e^{-\gamma{\left(t\right)}}\nonumber\\
    \rho_{46}{\left(t\right)}=\rho_{64}{\left(t\right)}=\frac{1}{6}e^{-4\gamma{\left(t\right)}}\nonumber\\  
     \rho_{24}{\left(t\right)}=\rho_{26}{\left(t\right)}=\rho_{35}{\left(t\right)}=\rho_{42}{\left(t\right)}=\rho_{53}{\left(t\right)}=\rho_{62}{\left(t\right)}=\frac{1}{6}e^{-2\gamma{\left(t\right)}}\nonumber\\
    \rho_{34}{\left(t\right)}=\rho_{43}{\left(t\right)}=\frac{1}{6}e^{-5\gamma{\left(t\right)}}.\nonumber
\end{gather}

\subsection{Independent random telegraph noise}\label{Pure qubit-qutrti_RTN}
The elements of the evolved density matrix, when each subsystem of the hybrid qubit-qutrit system is independently subject to the classical random telegraph noise, can be obtained as
\begin{gather}
    \rho_{11}{\left(t\right)}=\rho_{22}{\left(t\right)}=\rho_{33}{\left(t\right)}=\rho_{44}{\left(t\right)}=\rho_{55}{\left(t\right)}=\rho_{66}{\left(t\right)}=\frac{1}{6}\nonumber\\
        \rho_{12}{\left(t\right)}=\rho_{21}^{*}{\left(t\right)}=\frac{1}{6}e^{i\phi}D_{1}{\left(\tau\right)} \nonumber\\ 
        \rho_{23}{\left(t\right)}=\rho_{32}{\left(t\right)}=\rho_{45}{\left(t\right)}=\rho_{54}{\left(t\right)}=\rho_{56}{\left(t\right)}=\rho_{65}{\left(t\right)}=\frac{1}{6}D_{1}{\left(\tau\right)}\nonumber\\
        \rho_{13}{\left(t\right)}=\rho_{14}{\left(t\right)}=\rho_{31}^{*}{\left(t\right)}=\rho_{41}^{*}{\left(t\right)}=\frac{1}{6}e^{i\phi}D_{2}{\left(\tau\right)}\nonumber\\
        \rho_{25}{\left(t\right)}=\rho_{36}{\left(t\right)}=\rho_{46}{\left(t\right)}=\rho_{52}{\left(t\right)}=\rho_{63}{\left(t\right)}=\rho_{64}{\left(t\right)}=\frac{1}{6}D_{2}{\left(\tau\right)}\\
        \rho_{15}{\left(t\right)}=\rho_{51}^{*}{\left(t\right)}=\frac{1}{6}e^{i\phi}D_{2}{\left(\tau\right)}D_{1}{\left(\tau\right)}\nonumber\\
        \rho_{24}{\left(t\right)}=\rho_{26}{\left(t\right)}=\rho_{35}{\left(t\right)}=\rho_{42}{\left(t\right)}=\rho_{53}{\left(t\right)}=\rho_{62}{\left(t\right)}=\frac{1}{6}D_{2}{\left(\tau\right)}D_{1}{\left(\tau\right)}\nonumber \\
        \rho_{16}{\left(t\right)}=\rho_{61}^{*}{\left(t\right)}=\frac{1}{6}e^{i\phi}D_{2}^{2}{\left(\tau\right)}\nonumber\\
        \rho_{34}{\left(t\right)}=\rho_{43}{\left(t\right)}=\frac{1}{6}D_{2}^{2}{\left(\tau\right)}.\nonumber
\end{gather}

\subsection{Common random telegraph noise}\label{pure_qubit_qutrit_common_rtn}
The elements of the evolved  density matrix, when the qubit and qutrit  are subject to a common RTN source, are given by
\begin{gather}
    \rho_{11}{\left(t\right)}=\rho_{22}{\left(t\right)}=\rho_{33}{\left(t\right)}=\rho_{44}{\left(t\right)}=\rho_{55}{\left(t\right)}=\rho_{66}{\left(t\right)}=\frac{1}{6}\nonumber\\
        \rho_{12}{\left(t\right)}=\rho_{21}^{*}{\left(t\right)}=\frac{1}{6}e^{i\phi}D_{1}{\left(\tau\right)} \nonumber\\ \rho_{23}{\left(t\right)}=\rho_{32}{\left(t\right)}=\rho_{24}{\left(t\right)}=\rho_{42}{\left(t\right)}=\rho_{35}{\left(t\right)}=\rho_{53}{\left(t\right)}=\nonumber\\
   \rho_{45}{\left(t\right)}=\rho_{54}{\left(t\right)}=\rho_{56}{\left(t\right)}=\rho_{65}{\left(t\right)}=\frac{1}{6}D_{1}{\left(\tau\right)}\nonumber\\
        \rho_{13}{\left(t\right)}=\rho_{14}{\left(t\right)}=\rho_{31}^{*}{\left(t\right)}=\rho_{41}^{*}{\left(t\right)}=\frac{1}{6}e^{i\phi}D_{2}{\left(\tau\right)}\nonumber\\
        \rho_{25}{\left(t\right)}=\rho_{36}{\left(t\right)}=\rho_{46}{\left(t\right)}=\rho_{52}{\left(t\right)}=\rho_{63}{\left(t\right)}=\rho_{64}{\left(t\right)}=\frac{1}{6}D_{2}{\left(\tau\right)}\\
        \rho_{15}{\left(t\right)}=\rho_{51}^{*}{\left(t\right)}=\frac{1}{6}e^{i\phi}D_{3}{\left(\tau\right)}\nonumber\\
        \rho_{24}{\left(t\right)}=\rho_{26}{\left(t\right)}=\rho_{35}{\left(t\right)}=\rho_{42}{\left(t\right)}=\rho_{53}{\left(t\right)}=\rho_{62}{\left(t\right)}=\frac{1}{6}D_{2}{\left(\tau\right)}D_{1}{\left(\tau\right)}
        \nonumber\\
        \rho_{16}{\left(t\right)}=\rho_{61}^{*}{\left(t\right)}=\frac{1}{6}e^{i\phi}D_{4}{\left(\tau\right)}\nonumber\\
        \rho_{34}{\left(t\right)}=\rho_{43}{\left(t\right)}=\frac{1}{6}.\nonumber
\end{gather}

\subsection{Composite classical-quantum environments}\label{Pure hybrid qubit-qutrit subject composite environments}
The elements of the evolved  density matrix, when  the qubit and qutrit   are independently subject to, respectively, random telegraph noise channel and squeezed vacuum reservoirs, can be obtained as
\begin{gather}
    \rho_{11}{\left(t\right)}=\rho_{22}{\left(t\right)}=\rho_{33}{\left(t\right)}=\rho_{44}{\left(t\right)}=\rho_{55}{\left(t\right)}=\rho_{66}{\left(t\right)}=\frac{1}{6}\nonumber\\
        \rho_{12}=\rho_{21}^{*}=\frac{1}{6}e^{i\phi}e^{-\gamma{\left(t\right)}}\nonumber\\ 
        \rho_{23}{\left(t\right)}=\rho_{32}{\left(t\right)}=\rho_{45}{\left(t\right)}=\rho_{54}{\left(t\right)}=\rho_{56}{\left(t\right)}=\rho_{65}{\left(t\right)}=\frac{1}{6}e^{-\gamma{\left(t\right)}}\nonumber\\
        \rho_{13}{\left(t\right)}=\rho_{31}^{*}{\left(t\right)}=\frac{1}{6}e^{i\phi}e^{-4\gamma{\left(t\right)}}\nonumber\\
        \rho_{14}{\left(t\right)}=\rho_{41}^{*}{\left(t\right)}=\frac{1}{6}e^{i\phi}D_{2}{\left(\tau\right)}\nonumber\\
        \rho_{15}{\left(t\right)}=\rho_{51}^{*}{\left(t\right)}=\frac{1}{6}e^{i\phi}D_{2}{\left(\tau\right)}e^{-\gamma{\left(t\right)}}\\
        \rho_{16}{\left(t\right)}=\rho_{61}^{*}{\left(t\right)}=\frac{1}{6}e^{i\phi}D_{2}{\left(\tau\right)}e^{-4\gamma{\left(t\right)}}\nonumber\\
        \rho_{25}{\left(t\right)}=\rho_{36}{\left(t\right)}=\rho_{52}{\left(t\right)}=\rho_{63}{\left(t\right)}=\frac{1}{6}D_{2}{\left(\tau\right)}\nonumber\\
        \rho_{24}{\left(t\right)}=\rho_{26}{\left(t\right)}=\rho_{35}{\left(t\right)}=\rho_{42}{\left(t\right)}=\rho_{53}{\left(t\right)}=\rho_{62}{\left(t\right)}=\frac{1}{6}D_{2}{\left(\tau\right)}e^{-\gamma{\left(t\right)}}\nonumber\\
        \rho_{34}{\left(t\right)}=\rho_{43}{\left(t\right)}=\frac{1}{6}D_{2}{\left(\tau\right)}e^{-4\gamma{\left(t\right)}}\nonumber\\
        \rho_{46}{\left(t\right)}=\rho_{64}{\left(t\right)}=\frac{1}{6}e^{-4\gamma{\left(t\right)}}.\nonumber
\end{gather}

\section{Mixed hybrid qubit-qutrit evolved density matrix}\label{Mixed qubit-qutrit}
This appendix presents the elements of the evolved density matrix of hybrid qubit-qutrit system, starting from the initial mixed state of Eq.~(\ref{eq:Eq25}), in the presence of quantum and classical noises.

\subsection{Squeezed Vacuum reservoirs}
The elements of the evolved  density matrix, when each subsystem of the hybrid qubit-qutrit system is independently subject to a squeezed vacuum reservoir, are given by 
\begin{equation}\label{Density matrix S}
   \rho{(t)} =\left(
\begin{array}{cccccc}\frac{p}{2} & 0 & 0 & 0 & 0 & \frac{p}{2} \mathcal{F}\\
 0 & \frac{p}{2} & 0 & 0 & 0 & 0 \\
 0 & 0 & \frac{1-2 p}{2}& \frac{1-2 p}{2} \mathcal{F} & 0 & 0 \\
 0 & 0 & \frac{1-2 p}{2} \mathcal{F}  & \frac{1-2 p}{2} & 0 & 0 \\
 0 & 0 & 0 & 0 & \frac{p}{2} & 0 \\
 \frac{p}{2} \mathcal{F} & 0 & 0 & 0 & 0 & \frac{p}{2} \\
\end{array}
\right),
\end{equation} 
and the partial transpose with respect to the subsystem $A$ is
\begin{equation}\label{partial transpose}
\begin{split}
    {\left(\rho{(t)}^{AB}\right)}^{T_A}=\left(
\begin{array}{cccccc}\frac{p}{2} & 0 & 0 & 0 & 0 & \frac{1-2p}{2} \mathcal{F} \\
 0 & \frac{p}{2} & 0 & 0 & 0 & 0 \\
 0 & 0 & \frac{1-2 p}{2}& \frac{p}{2} \mathcal{F} & 0 & 0 \\
 0 & 0 & \frac{p}{2} \mathcal{F}& \frac{1-2 p}{2} & 0 & 0 \\
 0 & 0 & 0 & 0 & \frac{p}{2} & 0 \\
 \frac{1-2p}{2} \mathcal{F} & 0 & 0 & 0 & 0 & \frac{p}{2} \\
\end{array}
\right),
\end{split}
\end{equation} 
where the $\mathcal{F}=e^{-5\gamma{\left({(t)}\right)}}$.

\subsection{Independent random telegraph noise}
The elements of the evolved density matrix, when each subsystem of the hybrid qubit-qutrit system is  independently subject to the classical random telegraph noise, are given by Eq.~(\ref{Density matrix S}) with  $\mathcal{F}={D_{2}{(\tau)}}^{2}$.

\subsection{Common random telegraph noise}
The  evolved  density matrix, when the qubit and qutrit  are subject to a common RTN source, is given by
 \begin{equation}
   \rho{(t)} =\left(
\begin{array}{cccccc}\frac{p}{2} & 0 & 0 & 0 & 0 & \frac{p}{2} \mathcal{F}\\
 0 & \frac{p}{2} & 0 & 0 & 0 & 0 \\
 0 & 0 & \frac{1-2 p}{2}& \frac{1-2 p}{2}  & 0 & 0 \\
 0 & 0 & \frac{1-2 p}{2}  & \frac{1-2 p}{2} & 0 & 0 \\
 0 & 0 & 0 & 0 & \frac{p}{2} & 0 \\
 \frac{p}{2} \mathcal{F} & 0 & 0 & 0 & 0 & \frac{p}{2} \\
\end{array}
\right),
\end{equation} 
 where $\mathcal{F}=D_{4}{(\tau)}$.

\subsection{Composite classical-quantum environments}\label{APPB4}
The elements of the evolved  density matrix, when the qubit and qutrit  are independently subject to, respectively, random telegraph noise channel and squeezed vacuum reservoirs, are given by Eq.~(\ref{Density matrix S})
with $\mathcal{F}=D_{2}{(\tau)}e^{-4\gamma{(t)}}$.
%%%%%%%%%%%%%%%%%%%%%%%%%%%%%%%%%%%%%%%%%%

\bibliography{revisionHSS}% Produces the bibliography via BibTeX.
	
\end{document}